\documentclass[prd,preprint,superscriptaddress,showpacs,byrevtex]{revtex4}
\usepackage{epsfig}

\newcommand{\be}{\begin{equation}}
\newcommand{\ee}{\end{equation}}
\newcommand{\bea}{\begin{eqnarray}}
\newcommand{\eea}{\end{eqnarray}}
\def\simgt{\rlap{\lower 3.5 pt \hbox{$\mathchar \sim$}} \raise 1pt
Ê \hbox {$>$}}

\begin{document}

\title{Dark Matter Production at LHC from Black Hole Remnants }

\author{Gouranga C. Nayak} \email{nayak@physics.arizona.edu}
\affiliation{ Department of Physics, University of Arizona, Tucson, AZ 85721 USA
}


\begin{abstract}

We study dark matter production at CERN LHC from black hole remnants (BHR). We find that the
typical mass of these BHR at LHC is $\sim$ 5-10 TeV which is heavier than other dark matter
candidates such as: axion, axino, neutralino etc. We propose the detection of this dark matter
via single jet production in the process $pp \rightarrow $ jet +BHR(dark matter) at CERN LHC.
We find that for zero impact parameter partonic collisions, the monojet cross section is not
negligible in comparison to the standard model background and is much higher than the other
dark matter scenarios studied so far. We also find that $\frac{d\sigma}{dp_T}$ of jet production
in this process increases as $p_T$ increases, whereas in all other dark matter scenarios the
$\frac{d\sigma}{dp_T}$ decreases at CERN LHC. This may provide an useful signature for dark
matter detection at LHC. However, we find that when the impact parameter
dependent effect of inelasticity is included, the monojet cross section from the above process
becomes much smaller than the standard model background and may not be detectable at LHC.

\end{abstract}
\pacs{PACS: 95.35.+d; 04.50.Gh; 04.70.Dy; 13.85.-t } %
\maketitle

\newpage

By now it is confirmed that dark matter exists and it consists of a large fraction
of the energy density of the universe ($\sim$ 25 percent) \cite{dark} while dark energy
consists of $\sim$ 70 percent. The energy density of the non-baryonic dark matter in the
universe is known to be \cite{dens}
\bea
\Omega_{DM}h^2 = 0.112 \pm 0.009
\label{odark}
\eea
where $\Omega_{DM}$ is the energy density in units of the critical density
and $h\sim 0.71$ is the normalized Hubble parameter. Since the visible matter consists
of only $\sim$ 5 percent of the matter of the universe, the laws of
physics or laws of gravity as we know today may not be sufficient to explain the
dark matter and dark energy content of the universe.

One of the challenge we face today is to identify the non-baryonic
weakly interacting massive particle (WIMP) or WIMP-like particle which
consists of dark matter \cite{wimp}. Identification of this WIMP or WIMP-like
dark matter candidate is one of the outstanding questions in basic science today.
At present the possible proposals include: axion, axino,
neutralino, gravitino and black hole remnants etc. \cite{all}.
Black hole remnants as a source of dark matter is studied in various
inflation models in \cite{all,steinhard,chen}. These black hole remnants are
from black holes which were produced due to the density perturbations
in the early universe during inflation.

An exciting possibility is that black hole remnants (BHR) that make up some or all
of dark matter may be produced at high energy colliders such as large hadron
colliders (LHC) at CERN. Such prospects are particularly promising because both
ATLAS and CMS detectors at LHC will search for black holes. In this paper we study
dark matter production from black hole remnants at CERN LHC.

The Schwarzschild radius of $d~(=n+4)$ dimensional black hole is given by
\bea
R_{BH}=w_n \frac{1}{M_P} (\frac{M_{BH}}{M_P})^{\frac{1}{n+1}},~~~~~~~~~~w_n=(\frac{16 \pi}{(n+2)\Omega_{n+3}})^{\frac{1}{n+1}},
\label{rs}
\eea
where $M_{BH}$ is the black hole mass and $M_P$ is the Planck mass $\sim$ TeV at LHC
\cite{folks}. The Hawking temperature of the black hole becomes
\bea
T_{BH}= \frac{n+1}{4\pi R_{BH}}.
\label{tbh}
\eea

Once black hole is produced at LHC it will emit particles due to Hawking radiation
\cite{pp}. However, in the absence of a theory of quantum gravity
it is not clear what happens to black hole radiation when its mass approaches
Planck mass. It is commonly believed that quantum gravity implies the existence
of a minimum length \cite{garay} which leads to a modification of the quantum
mechanical uncertainty principle
\be
\Delta x \ge \frac{\hbar}{\Delta p} [1+(\alpha' L_P \frac{\Delta p}{\hbar})^2]
\label{dx}
\ee
where $L_P$ is the Planck length and $\alpha'$ is a dimensionless constant $\sim 1$
which depends on the details of the quantum gravity theory. The generalized
uncertainty principle (GUP), equation (\ref{dx}), can be derived in the context of
non-commutative quantum mechanics \cite{maggiore}, string theory \cite{amati} or
from minimum length considerations \cite{maggiore1}.

If we implement GUP and demand that the position uncertainty $\Delta x$ of the produced
particle from the black hole is of the order of Schwarzschild radius, then the
modified temperature of the black hole becomes \cite{chen,marco}
\bea
T_{BHR}=2T_{BH}[1+\sqrt{1-\frac{1}{w_n^2 (\frac{M_{BH}}{M_P})^{\frac{2}{n+1}}}}]^{-1}.
\label{tbhr}
\eea
The black hole temperature is undefined for $M_{BH} < M_{min}$ where
\bea
M_{min}= \frac{n+2}{8 \Gamma[\frac{n+3}{2}]} \pi^{\frac{n+1}{2}} M_P.
\label{mmin}
\eea
Black holes with mass less than $M_{min}$ do not exist, since their horizon radius
would fall below the minimum allowed length. Hence Hawking evaporation must stop
once the black hole mass reaches $M_{min}$. This creates a black hole remnant
of mass $M_{min}$ which is of $\sim$ TeV at LHC. Since this black hole remnant is
weakly interacting and heavy, it is a possible candidate for dark matter at LHC
\cite{steinhard,chen}.

Since the dark matter is weakly interacting it can not be directly detected at LHC.
For this purpose we will study dark matter production from black hole remnants (BHR) at
LHC in the process $pp \rightarrow $ jet + BHR(dark matter). We propose indirect detection
of dark matter via single jet measurement in the above process $pp \rightarrow $ jet
+ BHR(dark matter) at LHC.
The emission rate $\frac{dN}{dt}$ \cite{kribs} for jet production with momentum/energy $E =|\vec p|$ from a black hole,
which becomes a black hole remnant of mass $M_{min}$ after time $t_f$, is given by
\be
\frac{dN}{d^3p }=\int_0^{t_f}
\frac{c_s \sigma_s}{32\pi^3}\frac{dt}{(e^{\frac{E}{T_{BHR}}} \pm 1)}\,,
\label{thermal}
\ee
where $\sigma_s$ is the $d-$dimensional grey body factor \cite{greybody},
$T_{BHR}$ is the GUP implemented black hole temperature as given by eq. (\ref{tbhr}),
$t_f$ is the decay time \cite{marco} and $c_s$ is the multiplicity factor. $\pm$
is for quark and gluon jets respectively.

This result in Eq. ({\ref{thermal}}) is for jet production from a single
black hole of temperature $T_{BHR}$ (with a black hole remnant of mass $M_{min}$).
To obtain total jet cross section from this process we need to multiply
the number of jets produced from a single black hole with the total black hole
production cross section in pp collisions at LHC.

The black hole production cross section in pp collisions at $\sqrt{s}$= 14 TeV at
LHC is given by \cite{pp},
\bea
\sigma_{BH}^{pp \rightarrow BH }
= {\sum}_{ij}~
\int_{\tau}^1 dx_i \int_{\tau/x_i}^1 dx_j f_{i/p}(x_i, Q^2) \times f_{j/p}(x_j, Q^2)
\hat{\sigma}^{ij \rightarrow BH }(\hat s) ~\delta(x_i x_j -M_{BH}^2/s).
\label{bkt}
\eea
In this expression $\hat{\sigma}^{ab \rightarrow BH }(\hat s) = \pi R_{BH}^2$
is the black hole production cross section in partonic collisions at zero impact
parameter, $x_i (x_j)$ is the longitudinal momentum fraction of the parton inside
the proton at LHC and $\tau=M_{BH}^2/s$. Energy-momentum conservation implies
${\hat s} =x_ix_j s=M_{BH}^2$. We use $Q = \frac{1}{R_{BH}}$ as the factorization
scale at which the parton distribution functions are measured. ${\sum}_{ij}$
represents the sum over all partonic contributions where $i,j=q, {\bar q}, g$.

The above formula, eq. (\ref{bkt}), is valid for zero impact parameter partonic
collisions. To include the impact parameter dependent effect of inelasticity,
we adopt the impact parameter $b$ weighted average of the inelasticity used in
\cite{randall}
\bea
\sigma_{BH}^{pp \rightarrow BH }
= {\sum}_{ij}~\int_0^1 2z ~dz~\int_{\frac{(x_{\rm min} M_P)^2}{y^2(z)s}}^1 du \int_u^1 \frac{dv}{v}
f_{i/p}(v, Q^2) \times f_{j/p}(u/v, Q^2) \hat{\sigma}^{ij \rightarrow BH }(M_{BH}=\sqrt{us}) \nonumber \\
\label{bktin}
\eea
where $z=b/b_{max}$. The partonic level cross section is given by \cite{18}
\bea
\hat{\sigma}^{ij \rightarrow BH }(M_{BH}=\sqrt{us})= F(n) \pi R_S^2
\label{hatsig}
\eea
where
\bea
R_S = \frac{1}{M_P} [\frac{2^n \pi^{\frac{n-3}{2}}\Gamma[\frac{n+3}{2}]}{n+2}\frac{ \sqrt{us}}{M_P}]^{\frac{1}{n+1}}.
\label{rss}
\eea
The inelasticity parameter $y(z)$ and the cross section correction factor $F(n)$ are taken from \cite{26}.
We use the factorization scale $Q =\frac{1}{R_S}$ at which the parton distribution functions are measured.
$x_{\rm min}=\frac{M_{\rm BH}^{\rm min}}{M_P}$, where $M_{\rm BH}^{\rm min}$ is the smallest black hole
mass for which we trust semi-classical calculation.

The total jet production cross section in the process $pp \rightarrow $ jet +BHR(dark matter) at LHC is
then given by
\be
\sigma = N \times \sigma_{BH}
\label{mbhrdk}
\ee
where $\sigma_{\rm BH}$ is given by eq. (\ref{bkt}). To obtain $p_T$ distribution we use
$d^3p~=~2\pi ~dp_T ~p_T^2 ~dy ~{\rm cosh}y $ in eq. (\ref{thermal}). $y$ is the rapidity.

In our calculation we use CTEQ6M parton distribution functions inside the proton
\cite{cteq}. The number of extra dimensions is chosen to be $n=6$ so that we do not
rule out the possibility of Planck mass $M_P$ = 1 TeV \cite{planck}. Since initial
mass of the black hole must be greater than the Planck mass we choose
$M_i^{BH}$ = 5 $M_P$ in our calculation. It can be seen from eq. (\ref{mmin}) that
the black hole remnant mass $M_{min}$ does not depend on the black hole mass but
depends on the Planck mass and number of extra dimensions. We find that the typical
black hole remnant mass $M_{min}$ = 4.7 TeV for $M_P$ = 1 TeV and $M_{min}$ = 9.7 TeV
for $M_P$ = 2 TeV at LHC.

For a comparison we list here the lower limits on the Planck mass $M_P$ by various collider experiments.
The current limits from LEP2, CDF (run II) and D0 (run II) are as follows. The LEP2  analysis has set a lower
limit on the Planck mass $M_P^{\rm min}$=1.69 TeV by using graviton production \cite{lep}. Search for
large extra dimensions in the production of jets and missing transverse energy at CDF gives $M_P^{\rm min}$=0.83 TeV for n=6
to $M_P^{\rm min}$=1.18 TeV for n=2 \cite{cdf1}, where $n$ is the number of extra dimensions. The search for
large extra dimensions in final states containing one photon or jet and large missing transverse energy
at CDF gives $M_P^{\rm min}$=0.94 TeV for n=6 to $M_P^{\rm min}$=1.4 TeV for n=2 \cite{cdf2}.
Dielectron and diphoton measurements at D0 gives $M_P^{\rm min}$=1.3 TeV for n=7 to $M_P^{\rm min}$=2.1 TeV for n=2 \cite{d01}.
Search for large extra dimensions via single photon plus missing energy at D0 sets the limit $M_P^{\rm min}$=0.778 TeV
for n=8 to $M_P^{\rm min}$=0.884 TeV for n=2 \cite{d02}.

In Fig. 1 we present the monojet cross section, in the process $pp \rightarrow $
jet +BHR(dark matter), as a function of initial
black hole mass at CERN LHC. This result is for zero impact parameter partonic
collisions. The solid line is for Planck mass 1 TeV and the
dashed line is for Planck mass 2 TeV. It can be seen that for Planck mass
1 TeV and initial black hole mass 5 TeV the monojet cross section, in the
process $pp \rightarrow $ jet +BHR(dark matter), is
38.5 (pb). This value is much higher than the cross section 18.6 (fb) obtained
in other dark matter scenario with dark matter mass $\sim$ 100 GeV \cite{feng}.
In our case the dark matter mass (BHR mass) is 4.7 TeV which is much heavier
than 100 GeV dark matter mass used in \cite{feng}.

\begin{figure}[htb]
\vspace{2pt}
\centering{{\epsfig{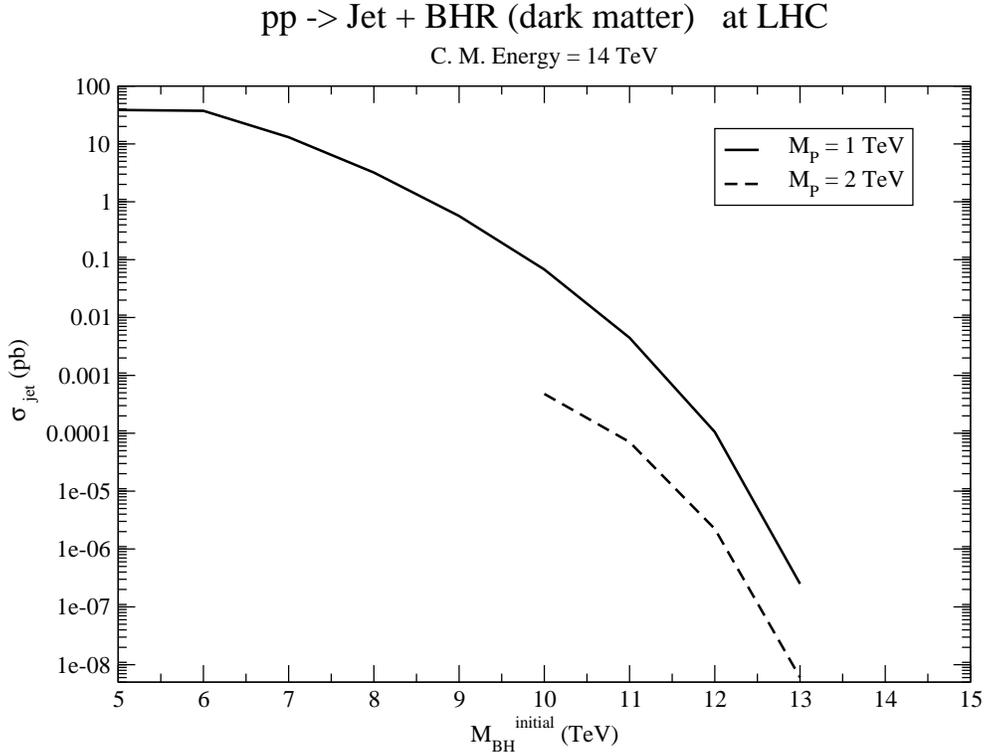}}}
\caption{ Total cross section for monojet production in the process
$pp \rightarrow $ jet + BHR (dark matter) at LHC at $\sqrt{s}$ = 14 TeV.}
\label{fig1}
\end{figure}

This is very exciting because we have found a heavier dark matter candidate at LHC
with larger cross section. This is due to the fact the temperature of a typical black
hole formed at LHC $\sim$ TeV. Hence jets produced from black holes at such high temperature
is large. On the other hand in other dark matter scenarios the jet plus dark matter
production is via direct parton collisions and hence the cross section is small.
Also unlike \cite{feng} our dark matter signal is not
negligible in comparison to the standard model background. A typical standard model background
is $\sim$ 130 pb for $p_T^{min}=100$ and GeV and 1300 pb for $p_T^{min}=30$ GeV.
In our case the cross section is $\sim$ 40 pb whereas in case of \cite{feng} the cross section
is 18.6 fb.

\begin{figure}[htb]
\vspace{2pt}
\centering{{\epsfig{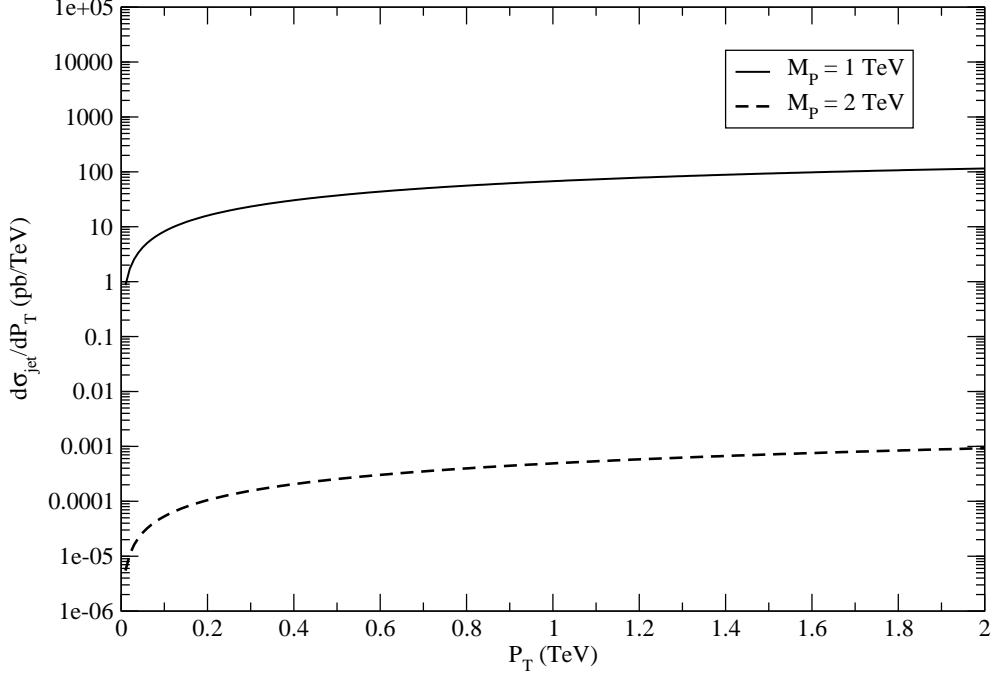}}}
\caption{$p_T$ differential cross section for monojet production in the process
$pp \rightarrow $ jet + BHR (dark matter) at LHC at $\sqrt{s}$ = 14 TeV.}
\label{fig2}
\end{figure}

In Fig. 2 we present the $p_T$ distribution of the jet production cross section,
in the process $pp \rightarrow$ jet +BHR(dark matter), at CERN LHC at $\sqrt{s}$
= 14 TeV. This result is for zero impact parameter partonic collisions.
The solid line is for Planck mass equals to 1 TeV and the dashed line is
for Planck mass equal to 2 TeV. It can be seen that $\frac{d\sigma}{dp_T}$ of
jet, from the process $pp \rightarrow$ jet +BHR(dark matter), increases as $p_T$
increases. This is in contrast to all other dark matter scenarios where $\frac{d\sigma}{dp_T}$
decreases as $p_T$ increases. This is also in contrast to all standard model processes where
$\frac{d\sigma}{dp_T}$ decreases as $p_T$ increases.

This is explained in detail in \cite{smith} and can be understood as follows. From the emission
rate $\frac{dN}{dt}$ in eq. (\ref{thermal}) we find
\be
\frac{dN}{dp_T }=2\pi p_T^2 \int dy~{\rm cosh}y~ \int_0^{t_f}
\frac{c_s \sigma_s}{32\pi^3}\frac{dt}{(e^{\frac{p_T~{\rm cosh}y}{T_{BHR}}} \pm 1)}\,.
\label{thermalfinal}
\ee
Since the temperature of the black hole remnant
$T_{BHR}\sim $ 1-2 TeV at LHC, the thermal distribution $\frac{1}{(e^{\frac{p_T~{\rm cosh}y}{T_{BHR}}} \pm 1)}$
remains almost flat with respect to $p_T$ as long as $p_T$ is not much larger than $T_{BHR}$. Hence the increase of
$\frac{d\sigma}{dp_T}$ as $p_T$ increases comes from the increase in the transverse momentum phase space factor $p_T^2$
as can be seen from eq. (\ref{thermalfinal}). For very large value of $p_T>>$ 2 TeV, the $\frac{d\sigma}{dp_T}$
will of course start decreasing. Hence the increase of  $\frac{d\sigma}{dp_T}$ as $p_T$ increases may provide
an unique signal for dark matter detection from black hole remnants at the CERN LHC.

In Fig. 3 we present the results which include the impact parameter
dependent effect of inelasticity in the cross section (see eq. (\ref{bktin})). We present
the monojet cross section, in the process $pp \rightarrow $
jet +BHR(dark matter), as a function of $x_{min}$ at CERN LHC. The solid line is for Planck mass 1 TeV
and the dashed line is for Planck mass 1.5 TeV. The monojet cross section is very small for $M_P$= 2 TeV
and hence we do not report it.
It can be seen that for Planck mass equal to 1 TeV and $x_{\rm min}$ equals to 5, the monojet cross
section, in the process $pp \rightarrow $ jet +BHR(dark matter), is 10 (fb) which is much smaller
than the zero impact parameter case (see Fig. 1). Hence when the impact parameter weighted average of the
inelasticity is included, the monojet cross section becomes much smaller than the standard model background
and may not be detectable at LHC.

\begin{figure}[htb]
\vspace{2pt}
\centering{{\epsfig{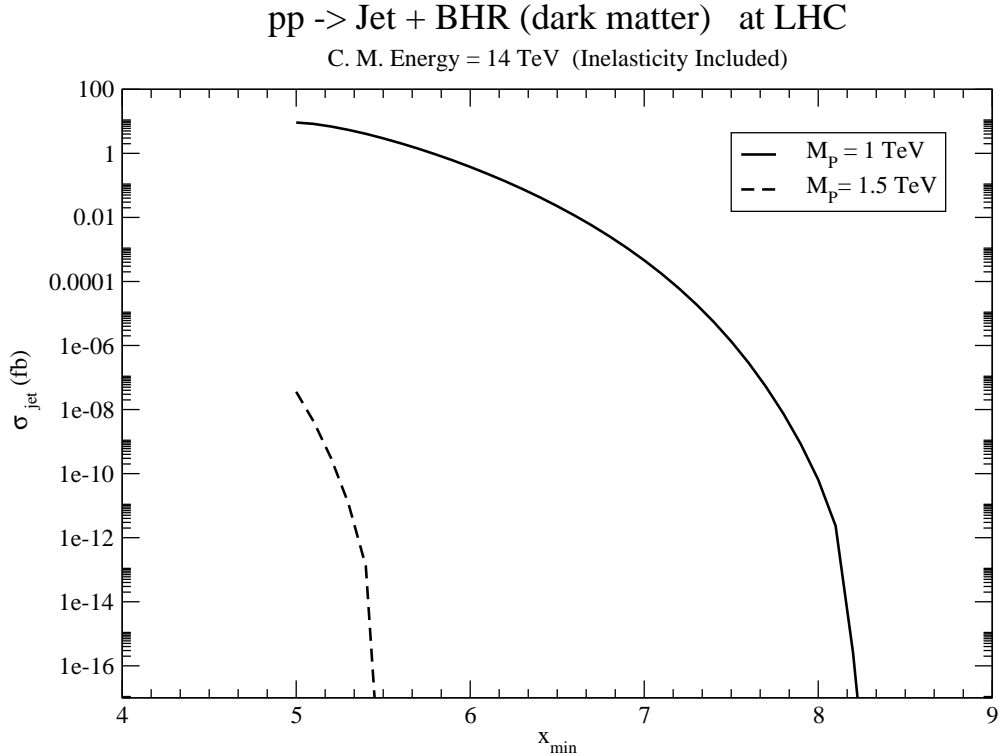}}}
\caption{ Total cross section for monojet production in the process
$pp \rightarrow $ jet + BHR (dark matter) at LHC at $\sqrt{s}$ = 14 TeV which
includes the effect of inelasticity.}
\label{fig3}
\end{figure}

In Fig. 4 we present the $p_T$ distribution of the cross section which include the impact parameter
dependent effect of inelasticity (see eq. (\ref{bktin})).
We use $x_{\rm min}$=5 in our calculation. The solid line is for Planck mass equals
to 1 TeV and the dashed line is for Planck mass equal to 1.5 TeV. The monojet cross section is very small
for $M_P$ = 2 TeV and hence we do not report it. It can be seen that $\frac{d\sigma}{dp_T}$
of jet, from the process $pp \rightarrow$ jet +BHR(dark matter), increases as $p_T$ increases. However, this
cross section is much smaller than the standard model background and may not be detectable at LHC. Only for
zero impact parameter partonic collisions, the cross section becomes comparable to the standard model predictions
(see Fig.2).

Finally we make some comments on the energy loss from a black hole to become a black hole remnant
and the TeV scale jets. For $M_P$ = 1 TeV and $M_{BH}$= 5 TeV, the mass of the black hole remnant is
$M_{BHR}$=4.7 TeV. Similarly for $M_P$ = 2 TeV and $M_{BH}$= 10 TeV, the mass of the black hole remnant is
$M_{BHR}$=9.7 TeV. Hence in both the cases the energy loss from a black hole to become a black hole remnant
is 300 GeV. One might wonder how can one compute high $p_T$ ($\sim$ 2 TeV) jets from black hole remnants in
Figs. 2 and 4. This is due to very high temperature of the black hole remnants. For $M_P$ = 1 TeV, $M_{BH}$=
5 TeV and $M_{BHR}$=4.7 TeV the temperature of the black hole remnant is $T_{BHR}$ = 0.98 TeV which can be
easily checked from eqs. (\ref{rs}), (\ref{tbh}) and (\ref{tbhr}). For $M_P$ = 2 TeV, $M_{BH}$= 10 TeV and
$M_{BHR}$=9.7 TeV the temperature of the black hole remnant is $T_{BHR}$ = 1.96 TeV. Hence the high $p_T$ jets
in Figs. 2 and 4 are due to very high temperatures ($T_{BHR}\sim$ 1-2 TeV) of the black hole remnants.

\begin{figure}[htb]
\vspace{2pt}
\centering{{\epsfig{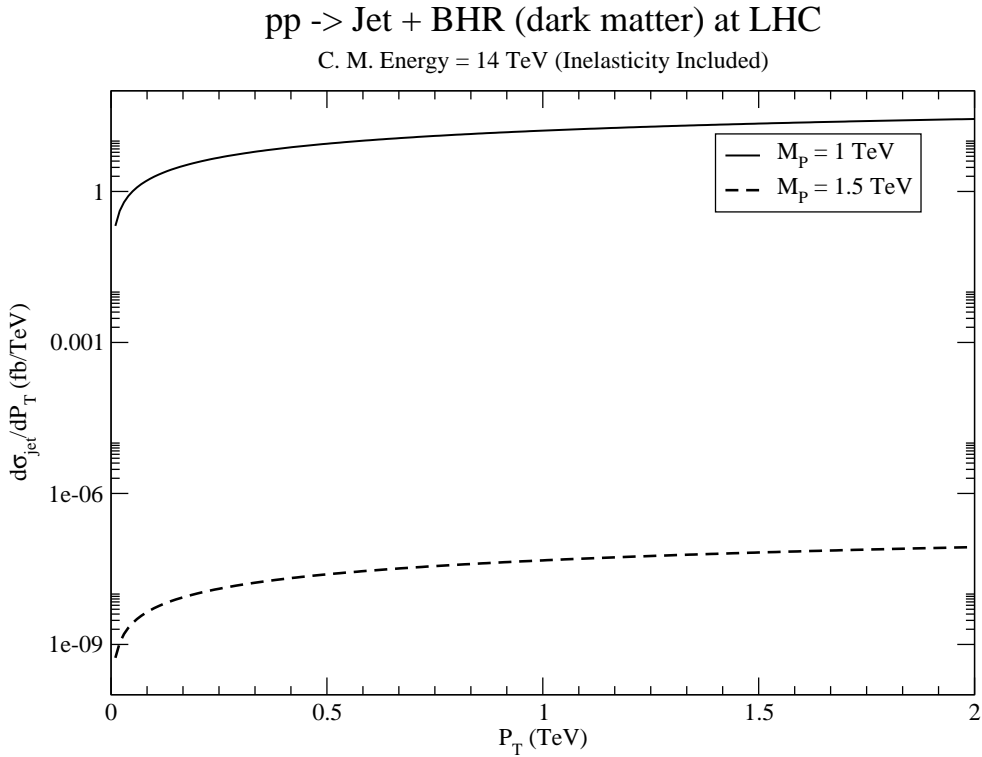}}}
\caption{$p_T$ differential cross section for monojet production in the process
$pp \rightarrow $ jet + BHR (dark matter) at LHC at $\sqrt{s}$ = 14 TeV which
includes the inelasticity.}
\label{fig4}
\end{figure}

To conclude, we have studied dark matter production at CERN LHC from black hole remnants (BHR).
We have found that the typical mass of these BHR at LHC is $\sim$ 5-10 TeV which is heavier
than other dark matter candidates such as: axion, axino, neutralino etc. We have proposed the
detection of this dark matter via single jet production in the process $pp \rightarrow $ jet
+BHR(dark matter) at CERN LHC. We have found that for zero impact parameter partonic collisions,
the monojet cross section is not negligible in comparison to the standard model background and is
much higher than the other dark matter scenarios studied so far. We have also found that
$\frac{d\sigma}{dp_T}$ of jet production in this process increases as $p_T$ increases, whereas in
all other dark matter scenarios the $\frac{d\sigma}{dp_T}$ decreases at CERN LHC. This may provide
an useful signature for dark matter detection at LHC. However, we have also shown that when the
impact parameter dependent effect of inelasticity is included, the monojet cross section from the above
process becomes much smaller than the standard model background and may not be detectable at LHC.

\acknowledgments

This work was supported in part by Department of Energy under
contracts DE-FG02-91ER40664, DE-FG02-04ER41319 and DE-FG02-04ER41298.

\end{document}